%% file: main_ieeetaes.tex
\documentclass[journal]{IEEEtran}

\usepackage{color,array,amsthm}
\usepackage{graphicx}

\ifCLASSINFOpdf
\else
   \usepackage[dvips]{graphicx}
\fi
\usepackage{url}
\usepackage{lipsum}
\usepackage{graphicx}
\usepackage[acronym]{glossaries}
\usepackage{nicefrac}
\usepackage{booktabs}
\usepackage{threeparttable}
\usepackage{subcaption}
\usepackage{xcolor}
\usepackage{pifont}
\usepackage{tcolorbox}
\usepackage[numbers,sort,square]{natbib}

\makeatletter
\newcommand{\removelatexerror}{\let\@latex@error\@gobble}
\makeatother

\usepackage[inline]{enumitem}

\usepackage{tikz}
\usepackage{pgfplots}
\usetikzlibrary{decorations.markings}
\usetikzlibrary{arrows}

\input{krkmath}

\definecolor{color1}{HTML}{648fff}
\definecolor{color2}{HTML}{785ef0}
\definecolor{color3}{HTML}{dc267f}
\definecolor{color4}{HTML}{fe6100}

\newacronym{sos}{SoS}{speed of sound}
\newacronym{mvdr}{MVDR}{minimum-variance distortionless response}
\newacronym{bra}{BRA}{Bayesian robust adaptive}

\makeatletter
\def\bstctlcite{\@ifnextchar[{\@bstctlcite}{\@bstctlcite[@auxout]}}
\def\@bstctlcite[#1]#2{%
 \@bsphack
 \@for\@citeb:=#2\do{%
 \edef\@citeb{\expandafter\@firstofone\@citeb}%
 \if@filesw\immediate\write\csname #1\endcsname{\string\citation{\@citeb}}\fi}%
 \@esphack}
\makeatother

\def\app#1#2{%
  \mathrel{%
    \setbox0=\hbox{$#1\sim$}%
    \setbox2=\hbox{%
      \rlap{\hbox{$#1\propto$}}%
      \lower1.1\ht0\box0%
    }%
    \raise0.25\ht2\box2%
  }%
}

\labelformat{subsection}{\thesection-#1}

\begin{document}
\bstctlcite{IEEEexample:BSTcontrol}

\title{Adaptive Bayesian Beamforming for Imaging by \\ Marginalizing the Speed of Sound} 

\author{Kyurae Kim, \IEEEmembership{Member, IEEE}, Simon Maskell, \IEEEmembership{Member, IEEE}, and Jason F. Ralph
\thanks{}
\thanks{K. Kim is with the University of Pennsylvania, Philadelphia, United States (e-mail: \url{kyrkim@seas.upenn.edu}).}
\thanks{S. Maskell and J. F. Ralph are with the University of Liverpool, Liverpool, United Kingdom (e-mail: \{S.Maskell, jfralph\}@liverpool.ac.uk).}
}





\maketitle

\begin{abstract}
    Imaging methods based on array signal processing often require a fixed propagation speed of the medium, also known as the~\acrfull*{sos} for methods based on acoustic signals.
    The resolution of the images formed using these methods is strongly affected by the assumed~\acrshort*{sos}, which, due to multipath, nonlinear propagation, and non-uniform mediums, is challenging to select.
    In this correspondence, we propose a Bayesian approach to marginalize the influence of the~\acrshort*{sos} on beamformers for imaging.
    In particular, we adapt a previously proposed Bayesian direction-of-arrival estimation model to the imaging setting, where we model the uncertainty of the~\acrshort*{sos} and then compute the posterior expectation of the popular minimum variance distortionless response beamformer (MVDR), resulting in a method we call \textit{Bayes MVDR}.
    Furthermore, we demonstrate that the marginal likelihood (\textit{ML}) of the model can be used for imaging.
    To solve the Bayesian integral efficiently, we use numerical Gauss-Hermite quadrature and apply our proposed approach to shallow water sonar imaging where multipath and nonlinear propagation are abundant.
    We compare Bayes MVDR and ML with the standard MVDR beamformer and demonstrate that its Bayesian counterpart achieves improved range and azimuthal resolution while effectively suppressing multipath artifacts.
\end{abstract}


\IEEEpeerreviewmaketitle

\section{Introduction}
\bstctlcite{IEEEexample:BSTcontrol}
\IEEEPARstart{B}{beamforming} algorithms \cite{knight_digital_1981,vanveen_beamforming_1988,vantrees_optimum_2002} for imaging require accurate physical modeling of the system to achieve high resolution.
It is often assumed that the signal propagates in a linear ``direct path'' with a globally fixed propagation speed of the medium.
This model is ``wrong but useful'', as George Box would comment, but discrepancies between the actual physics and the model result in lower image resolution and artifacts~\cite{holfort_investigation_2008}.
For example, sonar imaging in shallow waters suffers from multipath artifacts~\cite{blomberg_improving_2013}, 
ultrasound imaging has to deal with inhomogeneous media~\cite{pinton_sources_2011,perrot_you_2021}.
Since we focus on applications with complex acoustic propagation (such as sonar, medical ultrasound, and seismology), with some loss of generality, we will hereafter refer to the propagation speed as the \acrfull*{sos}.

\textcolor{black}{
Modeling the \acrshort*{sos} of sonar is particularly difficult since the propagation speed varies by depth, salinity, ocean temperature, and weather.
As such, modeling the \acrshort*{sos} profile~\citep[\S 9.1.2]{ainslie_principles_2010} is challenging, and the resulting profile will often be imperfect.
As such, for beamforming, selecting the right \acrshort*{sos} is not only difficult, but any fixed choice will often suffer from modeling imperfections.
}

Various approaches have been proposed to combat the issue of~\acrshort*{sos} selection.
Some authors have proposed to statistically estimate the propagation speed~\cite{stahli_bayesian_2021,ali_local_2022}, and some of these have been specifically proposed in the context of beamforming~\cite{perrot_you_2021}.
Although the method of Stahli \textit{et al.} \cite{stahli_bayesian_2021} provides pixel-wise estimates of the~\acrshort*{sos}, it requires \textit{a priori} segmentation of the target field, and its utility for beamforming has yet to be studied.

Meanwhile, adaptive beamformers, such as variants of the \acrfull*{mvdr} beamformer, which are purely statistical signal processing-based approaches,~\cite{vanveen_beamforming_1988,krim_two_1996} have been shown to be effective against \acrshort*{sos} mismatch~\cite{holfort_investigation_2008} and other inconveniences such as cross-channel interference, noise~\cite{shan_adaptive_1985}, and multipath~\cite{blomberg_improving_2013}.
\textcolor{black}{
However, these approaches do not directly try to \textit{correct} the chosen~\acrshort*{sos}.
In this work, we extend \acrshort*{mvdr} in a way that directly deals with the imperfections in the choice of~\acrshort*{sos} by modeling it as an uncertain quantity via probabilistic modeling.
In particular, our probabilistic model is inspired by the approach of Bell \textit{et al.} \cite{bell_bayesian_2000}.
By estimating the uncertainty of the~\acrshort*{sos}, we are able to \textit{marginalize} it, resulting in a marginalized beamformer achieving better resolution.
}

Bell \textit{et al.} \cite{bell_bayesian_2000} proposed what they call the~\acrfull*{bra} for direction-of-arrival (DoA) estimation.
In more detail, they adapt~\acrshort*{mvdr} by performing Bayesian inference of the optimal beam steering direction, where the likelihood uses the covariance of the received signal.
The posterior can be used to marginalize the steering direction used within the~\acrshort*{mvdr} beamformer~\cite{frost_algorithm_1972}.
The~\acrfull*{bra} beamformer has been further extended for multiple DoA estimation~\cite{chakrabarty_bayesian_2018} and passive sonar detection~\cite{yocom_bayesian_2011}, while the convergence of this method has been established by Lam and Singer~\cite{lam_bayesian_2006}.

When applying beamforming to imaging, each pixel in the image becomes a focal point.
Therefore, the DoA estimation framework of \citep{bell_bayesian_2000} cannot be applied directly.
However, in this correspondence, we demonstrate that the DoA estimation framework can be used for pixel-wise~\acrshort*{sos} estimation. 
Firstly, we estimate the pixel-wise~\acrshort*{sos} by adopting the likelihood of the \acrshort*{bra} beamformer.
Then, the~\acrshort*{sos} used to compute the \acrshort*{mvdr} beamformer is marginalized over the posterior for the~\acrshort*{sos} forming what we call ``Bayes MVDR.''
Our approach provides a data-driven way to handle uncertainties associated with the propagation model by expressing them as uncertainties in the \acrshort*{sos}.
Furthermore, by adopting a probabilistic model for the acquired signals, we can now compute the \textit{evidence}, or marginal likelihood, which can be used to detect the presence of signals in contrast to noise.
We demonstrate that this quantity, denoted ``ML'' in the experiments, can be used for imaging and can further suppress noise and multipath artifacts compared to Bayes MVDR.
For posterior inference, we employ Gauss-Hermite quadrature, which, unlike previous works in Bayesian DoA estimation, enables the proper use of priors with continuous support, such as Gaussians.

We evaluate our method by applying it to active sonar imaging.
We use the Bellhop propagation model~\cite{porter_gaussian_1987,bucker_simple_1994} to simulate point targets.
Compared to the MVDR beamformer~\cite{frost_algorithm_1972,shan_adaptive_1985,carlson_covariance_1988,asl_contrast_2011} popularly used in sonar imaging~\cite{stergiopoulos_implementation_1998,gerstoft_adaptive_2003,blomberg_adaptive_2012,austeng_use_2013,buskenes_optimized_2015,buskenes_lowcomplexity_2016,birkeneslonmo_improving_2020}, our method effectively suppresses multipath artifacts while achieving lower sidelobes and improved azimuth resolution.

To summarize, our contributions are as follows:
\begin{itemize}
    \item We propose a Bayesian model for estimating the speed of sound from the received signal on a pixel-by-pixel basis.
    
    \item We propose adaptive beamforming algorithms for imaging, Bayes MVDR and ML, which utilize the posterior of the \acrshort*{sos}.
    
    \item We propose the use of an efficient numerical quadrature scheme for computing the posterior.
\end{itemize}
We discuss our motivations and the setup in more detail in \cref{section:background}, where the signal model is described in \cref{section:signal_model}.
The probabilistic model is presented in \cref{section:probabilistic_model}, with the inference strategy in \cref{section:quadrature} and the resulting beamformer in \cref{section:beamformer}.
We then present experimental results in \cref{section:experiments} before concluding in \cref{section:conclusions}.

\section{Adaptive Bayesian Beamforming by Marginalizing of the Speed of Sound}
\subsection{Background and Motivation}\label{section:background}
\textit{
Conventional Beamforming.
}~
The conventional delay-and-sum (DAS,~\cite{vantrees_optimum_2002}) beamforming method, also known as backprojection, is described as
{%
\begin{align}
   y\left(p\right) 
   = \sum^{N}_{n=1} w_n \, x_n\big(t\,(p, c, n)\big) 
   = \vw^{\dagger} \vx^{\rm{d}}\left(p, c\right),
\end{align}
}%
where \(\dagger\) is the conjugate transpose, \(x_n\left(t\right)\) is the signal sample received by the \(n\)th sensor at time \(t\), \(p\) is the focal point (or image pixel), \(y\left(p\right)\) is the beamformed response of \(p\), \(c\) is the signal propagation speed of the medium, \(n = 1, \ldots, N\) is the sensor index, \(t\left(p, c, n\right)\) is the round-trip time for the response of \(p\) to reach the \(n\)th sensor, \(w_n\) is known as the fading weight or apodization weight, \(\vw\) is the fading weight vector, and \(\vx^{\rm{d}}\left(p, c\right)\) is the delayed signals in vector form.
The fading weights are often chosen independently of the received data to be off-the-shelf window functions, such as the Hann or Hamming windows.


The round-trip time \(t\left(p, c, n\right)\) is computed as
{%
\begin{align}
   t\left(p, c, n\right)
   =
   \frac{r\left(p, c, n\right)}{c}
   =
   \frac{r_{\mathrm{TX}}\left(p, c\right) + r_{\mathrm{RX}}\left(p, c, n\right)}{c},
    \label{eq:propagation_time}
\end{align}
}%
where \(r\left(p, c, n\right)\) is the round-trip distance from the source to the \(n\)th array sensor, \(r_{\mathrm{TX}}\) is the distance from source to the \(p\), \(r_{\mathrm{RX}}\) is the distance from \(p\) to the array sensor.
In this work, we consider a 2-dimensional imaging setting over the azimuth (\(x\)-axis) and range (\(y\)-axis).

\textit{
Adaptive Beamforming.~
}
While simple and effective, DAS tends to result in poor resolution and high sidelobes.
As a remedy, it is possible to \textit{adapt} the fading weight to the received data, a scheme often called \textit{adaptive} beamforming~\citep{vanveen_beamforming_1988}.
The most widely used approach is the celebrated minimum-variance distortionless response (MVDR) beamformer~\citep{frost_algorithm_1972}, also known as the Capon beamformer~\citep{capon_highresolution_1969}, which sets the fading weights as
{%
\begin{align}
    \vw\left(p,c\right) = \frac{
        \widehat{\mSigma}^{-1}\left(p, c\right) \mathbf{1}
    }{
        \mathbf{1}^{\top} \widehat{\mSigma}^{-1}\left(p, c\right) \mathbf{1}
    },\label{eq:mvdr}
\end{align}
}%
where \textcolor{black}{\(\mathbf{1} = {[1 \; \ldots \; 1]}^{\top}\) is a vector of ones}, \(\widehat{\mSigma}\) is the noise covariance of the signal at the focal point \(p\) given \(c\), which is estimated from the received data.
MVDR is known to greatly improve the resolution of the DAS beamformer while suppressing the sidelobes, provided that \(c\) has been set appropriately.

\textit{
Choosing the Speed of Sound.~
}
\cref{eq:propagation_time} assumes that the signal propagates linearly and at a constant speed \(c\), the~\acrshort*{sos}.
Choosing \(c\) is crucial to the resolution of the resulting image for both  conventional~\citep{perrot_you_2021} and adaptive beamformers~\cite{holfort_investigation_2008}.
We assert that the common practice of choosing a globally fixed~\acrshort*{sos} is suboptimal.
This is especially apparent in sonar applications where the propagation path is nonlinear and varies depending on oceanic parameters such as the \acrshort*{sos} profile~\cite{ainslie_principles_2010}.
Furthermore, active sonar suffers from \textit{multipath}~\citep{blomberg_adaptive_2013}, where a signal traveling between two points can take multiple paths of different lengths.
Under the direct-path model with a single fixed \acrshort*{sos}, multipath results in artifacts where the same target appears in multiple locations.
In this work, we avoid this problem by incorporating the \textit{uncertainty} of the \acrshort*{sos}, which filters out these multipath artifacts.

\subsection{Signal Model}\label{section:signal_model}
Before presenting our Bayesian beamforming approach, we discuss our signal model.
Let a signal be generated from a focal point \(p\).
The spectrum of this signal received by an \(N\)-sensor array is modeled as
\vspace{-0.02in}
{
\begin{align}
    \vx\left(\omega\right) = s\left(p, \omega\right) \va\left(p, c\right) + \vn\left(p, \omega\right)\label{eq:signal_model}
\end{align}
}%
where \(\omega\) is the frequency, \(s\) is the envelope of the original target response, \(\va \in \mathbb{C}^{N}\) is the steering vector defined as
\vspace{-0.02in}
{
\begin{align}
    \va\left(p, c\right) = 
    {\begin{bmatrix}\; 
        e^{\mathrm{j}\, \omega \, t\,\left(p,\, c,\, 1\right)} 
        &
        \ldots
        &
        e^{\mathrm{j}\, \omega \, t\,\left(p,\, c,\, N\right)}\;
    \end{bmatrix}}^{\dagger}
\end{align}\vspace{-0.02in}%
}%
for the time delay \(t\left(p, c, n\right)\), \(\mathrm{j}\) is the imaginary number, and \(\vn \in \mathbb{C}^{N}\) is the multivariate interference and noise vector.
The dependence on \(c\) signifies that the response of the same focal point \(p\) can take different paths, which are identified by \(c\).

The covariance of the received signal \(\mSigma_{\vx} \in \mathbb{C}^{N \times N}\) is
%
{
\begin{align}
    \mSigma_{\vx}\left(p, c\right) = \sigma^2_s\left(p\right) \; \va\left(p, c\right) \va^{\dagger}\left(p, c\right) + \mSigma_{\vn}\left(p\right),
    \label{eq:covariance_model}
\end{align}
}%
where \(\mSigma_{\vn} \in \mathbb{C}^{N \times N}\) is the noise (and interference) covariance matrix.
Note that if the data \(x_n\) is pre-delayed in the time domain by \(t\,\left(p,\, c,\, n\right)\), the steering vector is replaced by a vector of 1s.

\vspace{-0.03in}

\vspace{-0.1in}
\subsection{Probabilistic Model}\label{section:probabilistic_model}
\textit{
Likelihood.
}~
We now describe our probabilistic model used to infer the \acrshort*{sos}.
Our likelihood is similar to that used by the \acrshort*{bra} beamformer~\cite{bell_bayesian_2000} and is described as 
{%
\vspace{-0.02in}
\begin{align}
    &\ell_p\left(\vx_{1\text{:}K} \mid c\right) 
    = \prod^{K}_{l=1} \mathcal{CN}\left(\vx_k\left(p, c\right);\; \mathbf{0},\, \Sigma_{\vx}\left(c, p\right) \right)
    \\
    &\propto
    {\abs{\mSigma_{\vx}\left(p, c\right)}}^{-K}
    \exp\left( 
        -\sum^{K}_{l=1}
        \vx_k^{\dagger}\left(p, c\right) \mSigma^{-1}_{\vx}\left(p, c\right) 
        \vx_k\left(p, c\right)
    \right),\label{eq:likelihood}
\end{align}
}%
where \(\mathcal{CN}\) is the probability density function of a complex Gaussian and the data points \(\vx_{k}\left(p, c\right) \in \mathbb{C}^L\) 
are obtained by dividing the \(N\) sensor array into \(K = N - L + 1\) overlapping subarrays such that
\(
    \vx_k\left(p, c\right) = {\big[\,
        x_{k+1} \;
        x_{k+2} \;
        \ldots \;
        x_{k+L-1}
    \,\big]}^{\top}.
\)
Note that we have omitted the argument \(t\left(p, c, l\right)\) for clarity.
By using the subarray scheme~\cite{shan_adaptive_1985}, we avoid cancellation between coherent signals and obtain multiple samples without considering a temporal window.
It also guarantees that the estimated covariance matrix is full-rank under mild assumptions~\cite{shan_adaptive_1985}.

The determinant in \cref{eq:likelihood} penalizes covariances with large determinants, which implies that the signal is incoherent and has high variance, while the exponential term penalizes covariances that fail to explain the obtained data.
Therefore, this likelihood selects values of \(c\) that, after focusing on the point \(p\), gives rise to resulting observations that are
coherent and low variance.
Since the exact expression for \(\mSigma_{\vx}\) is unknown due to our lack of knowledge on \(\mSigma_{\vn}\) and \(\sigma^2_s\), approximation is required.


\textit{
Approximate Likelihood.
}~
From the definition of \(\mSigma_{\vx}\) in~\cref{eq:covariance_model}, Bell \textit{et al.}~\cite{bell_bayesian_2000} have shown that the likelihood can be represented as
{
\begin{align}
    &\ell_p\left(\vx_{1\text{:}K} \mid c\right) 
    \nonumber
    \;\propto
    {\left(1 + \sigma^2 \beta\left(c\right)\right)}^{-\textcolor{black}{K}}
    \nonumber
    \\
    &\quad \times
    \exp\left(
    \frac{
        \textcolor{black}{K}\,\sigma^2_s\,\beta^2\left(c\right)
    }{
        1 + \sigma^2 \beta\left(c\right)
    }
    \,
    \frac{
        \va^{\top}
        \mSigma_{\vn}^{-1}\left(\vc\right)
        \widehat{\mSigma}_{\vn}^{-1}\left(\vc\right)
        \mSigma_{\vn}^{-1}\left(\vc\right)
        \va
    }{
        \beta^2\left(c\right)
    }
    \right)\label{eq:likelihood_approx}
\end{align}
}%
where \(\beta\left(c\right) = \va^{\dagger} \mSigma_{\vn} \va\). Note that we have dropped the parameter \(p\) for clarity.
To avoid dealing with \(\mSigma_{\vn}\), Bell \textit{et al.} propose to further approximate the likelihood as
{
\begin{align}
    \ell_p\left(\vx_{1\text{:}K} \mid c\right) 
    \approx
    \widetilde{\ell}_p\left(\vx_{1\text{:}K} \mid c\right) 
    =
    C
    \exp\big(
       \textcolor{black}{K} \; \gamma \; P_{s}(c)
    \big), \label{eq:final_likelihood}
\end{align}
}%
where 
\(C > 0\) is some constant, \(P_{s}\) is the spectral estimate of the optimal output power \(\sigma_s^2\), and \(\gamma\) is a constant controlling the strength of the likelihood depending on the signal-to-noise (SNR) ratio.

The Capon~\cite{capon_highresolution_1969} spectral estimate of the optimal output power is defined as
{%
\begin{align}
    P_{s}\left(c\right) = {\left(\va^{\dagger} {\mSigma_{\vn}}^{-1}\left(c\right) \va\right)}^{-1}
    \label{eq:capon_spectral_estimate}
\end{align}
}%
and is the only point where the data enters our likelihood.
Therefore, accurately estimating \(P_{s}\) is crucial to the performance of our beamformer.
We further discuss this in~\cref{section:quadrature}.
Since \(\mSigma_{\vn}\left(c\right)\) is a nonlinear function of \(c\), our model is not linear.

The power strength constant is defined as
{%
\begin{align}
    \gamma 
    = 
    \frac{\textcolor{black}{K}}{\sigma_{n}^4} \,
    \frac{\textcolor{black}{K}\,\nicefrac{\sigma^2_s}{\sigma^2_n}}{1 + \textcolor{black}{K}\,\nicefrac{\sigma^2_s}{\sigma^2_n}}
    =
    \frac{\textcolor{black}{K}}{\mathsf{NL}^2} \,
        \frac{K \, \mathsf{SNR}}{1 + K \, \mathsf{SNR}},\label{eq:gamma}
    %
\end{align}
}%
where \(\mathsf{SNR} = \nicefrac{\sigma^2_s}{\sigma^2_n}\) is the signal-to-noise ratio and \(\mathsf{NL} = \sigma_n^2\) is the noise level.

\textit{
Power Strength \(\gamma\).
}~
While the original passive detection setting of Bell \textit{et al.} allowed the use of a fixed \(\mathsf{SNR}\) and \(\mathsf{NL}\), this is not the case for our active imaging setting.
In active imaging systems, the signal level (and thus the SNR) decreases with distance due to path loss.
To compensate for this, a time-varying gain (TVG) gain is applied proportionally to the round-trip distance \(r_{p}\).
This results in the ambient noise level \(\mathsf{NL}\) being amplified with distance.
We thus define a focal point dependent \(\gamma\left(p\right)\) comprised of
\begin{align}
    \mathsf{NL}\left(p\right) \texttt{[dB]} &= \mathsf{DR} - \mathsf{SNR}_0  + G_{\mathrm{TVG}}\left(r_p\right) \label{eq:nl} \\
    \mathsf{SNR}\left(p\right) \texttt{[dB]} &= \mathsf{SNR}_0 - \mathsf{PL}\left(p\right) \approx \mathsf{SNR}_0  - G_{\mathrm{TVG}}\left(r_p\right), \label{eq:snr} 
\end{align}
where \(\mathsf{DR}\) is the dynamic range, \(\mathsf{SNR}_0\) is the optimal SNR, \(\mathsf{PL}\left(p\right)\) and  \(G_{\mathrm{TVG}}\left(r_p\right)\) are the path loss and TVG for the focal point \(p\).
Note that \cref{eq:nl} assumes that the signal tightly fits the dynamic range while \cref{eq:snr} assumes that the TVG is set similarly to the path loss.

\vspace{-2ex}
\subsection{Bayesian Inference}\label{section:quadrature}
\textit{
Spectral Estimation.
}~
Estimating \(P_{s}\) is normally carried out by first substituting the empirical data covariance \(\widehat{\mSigma}_{\vx}\) for the noise covariance \(\mSigma_{\vn}\) such that \(P_{s} \approx {\left( \va^{\dagger} \widehat{\mSigma}_{\vx} \va \right)}^{-1}\).
For estimating the covariance, as discussed in~\cref{section:probabilistic_model}, we use subarray averaging~\cite{shan_adaptive_1985}, where the \(N\)-sensor array is divided into \(\textcolor{black}{K}\) subarrays.
This leads to an \(\textcolor{black}{K}\)-sample estimate of the covariance matrix, 
{
\begin{align}
    \widehat{\mSigma}_{\vx}\left(p, c\right) = \frac{1}{K} \sum^{K}_{k=1}
    \vx_{k}\left(p, c\right) \vx^{\dagger}_{k}\left(p, c\right).
\end{align}
}%
Without subarray averaging, cancellations of coherent signals result in a significant underestimation of the signal power.
To avoid this, we use the forward-backward averaged covariance~\cite{evans_application_1982,pillai_forward_1989,li_performance_1998} defined as
{%
\begin{align}
    \widehat{\mSigma}_{\vx,\,\text{FB}}\left(p, c\right)
    =
    \frac{1}{2}
    \left(
    \widehat{\mSigma}_{\vx}\left(p, c\right)
    + 
    \mJ \,{\widehat{\mSigma}_{\vx}^{\top}\left(p, c\right)}\, \mJ
    \right),
\end{align}
}%
where \(\mJ\) is the exchange matrix with ones only on the anti-diagonal, and diagonal loading~\cite{carlson_covariance_1988} 
{%
\begin{align}
    \widehat{\mSigma}_{\vx,\,\text{DL}}\left(p, c\right)
    =
    \widehat{\mSigma}_{\vx,\,\text{FB}}\left(p, c\right) + \epsilon\, \text{trace}\left\{\,\widehat{\mSigma}_{\vx,\,\text{FB}}\left(p, c\right)\,\right\} \, \mI
    \label{eq:covariance_diagonal_loaded}
\end{align}
}%
with \(\epsilon = 10^{-3} / \textcolor{black}{K}\) as recommended by Featherstone~\cite{featherstone_novel_1997}.
These modifications significantly improve the accuracy of the estimated power~\cite{asl_contrast_2011}, which is crucial for our approach.


\setcounter{figure}{2}

\setcounter{figure}{0}
\begin{figure*}
    \begin{subfigure}[b]{0.31\textwidth}
        \includegraphics[scale=0.85]{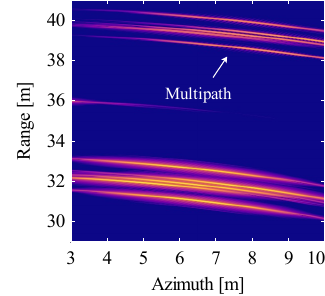}
        \caption{DAS}
    \end{subfigure}
    \begin{subfigure}[b]{0.31\textwidth}
        \includegraphics[scale=0.85]{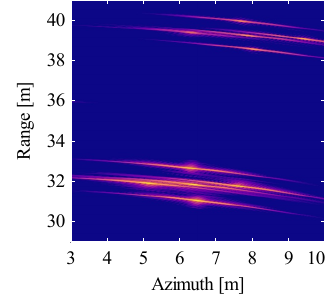}
        \caption{MVDR}
    \end{subfigure}
    \begin{subfigure}[b]{0.31\textwidth}
        \includegraphics[scale=0.85]{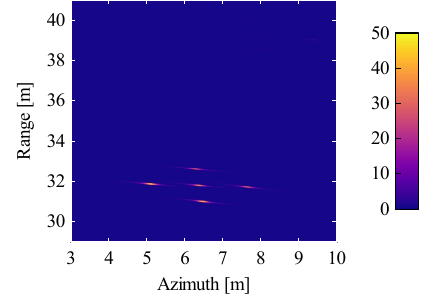}
        \caption{ML (\(Q=8\))}
    \end{subfigure}
    \\
    \begin{subfigure}[b]{0.31\textwidth}
        \includegraphics[scale=0.85]{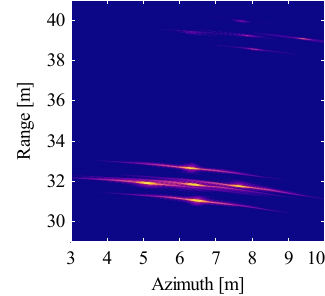}
        \caption{Bayes MVDR (\(Q\)=8)}
    \end{subfigure}
    \begin{subfigure}[b]{0.31\textwidth}
        \includegraphics[scale=0.85]{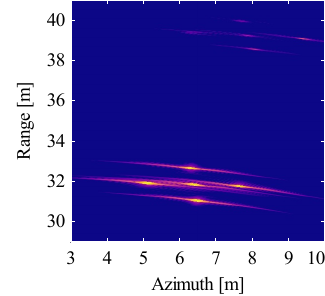}
        \caption{Bayes MVDR (\(Q\)=16)}
    \end{subfigure}
    \begin{subfigure}[b]{0.37\textwidth}
        \includegraphics[scale=0.85]{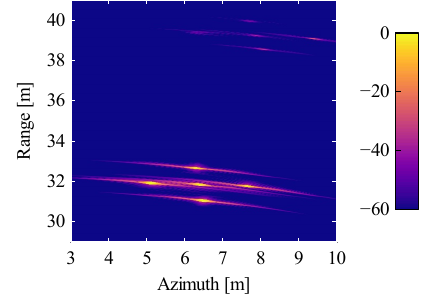}
        \caption{Bayes MVDR (\(Q\)=32)}
    \end{subfigure}
    \caption{
    \textbf{Images formed by the methods considered in this work.
    }
    The dynamic range of energy-based imaging methods is restricted to 0 dB to 60 dB (color bar in the second row), while the dynamic range of log ML was not restricted and is shown within the range of 0 nats to 50 nats (color bar in the first row).
    Compared to MVDR, the proposed schemes (Bayes MVDR and ML) achieve lower sidelobes and less pronounced multipath artifacts.
    }\label{fig:qualitative}
    \vspace{-2ex}
\end{figure*}

\textit{
Prior Distribution.
}~
For DoA estimation, it is natural to consider an uninformative prior on the steering angle.
Furthermore, using a discrete distribution simplifies the computation associated with performing Bayesian inference.
Thus, the previous works~\cite{yocom_bayesian_2011,chakrabarty_bayesian_2018} have considered setting a uniform discrete prior with a support of \([-\pi, \pi]\) on the steering angle.

For the speed of sound, we are in a slightly different position:
\begin{enumerate*}[label=(\roman*)]
    \item We have an informed guess about the optimal \acrshort*{sos},
    \item but do not know how much the true optimal \acrshort*{sos} will deviate from it.
\end{enumerate*}
Therefore, we consider a continuous informative prior distribution with ``soft'' boundaries.
We choose the prior to follow a normal distribution as \(\rho\left(c\right) = \mathcal{N}\left(c; \mu_c, \sigma^2_c\right)\) with mean \(\mu_c\) and standard deviation \(\sigma_c\). 

\textit{
Inference with Gauss-Hermite Quadrature.
}~
By choosing a Gaussian prior, we can use Gauss-Hermite quadrature~\cite{hildebrand_introduction_2003}.
This provides a principled approach to computing the marginalizing integral (discussed in \cref{section:beamformer}) without relying on arbitrary discretization as done in the previous works~\cite{yocom_bayesian_2011,chakrabarty_bayesian_2018}.
Given a computational budget of \(Q\) quadrature points, the posterior \(\pi\) is 
{
\begin{align}
    \pi_p\left(c \mid \vx_{1\text{:}K}\right)
    &=
    \frac{1}{Z\left(p\right)}
    \widetilde{\ell}_p\left( \vx_{1\text{:}K} \mid c \right) \, \rho\left(c\right)
    \nonumber
    \\
    &\propto
    \widetilde{\ell}_p\left(\vx_{1\text{:}K} \mid c \right)
    \,
    \exp\left(-\frac{1}{2} \frac{{\left(c - \mu_c\right)}^2}{\sigma_c^2}\right) 
    \nonumber
    \\
    &=
    \widetilde{\ell}_p\left(
     \vx_{1\text{:}K} \mid
    \sqrt{2} \, z \, \sigma_c + \mu_c \right)
    \,
    \mathrm{e}^{-z^2},
\end{align}
}%
where we have reparameterized as \(c = \sqrt{2} \, z \, \mu_c + \sigma_c\) and \(Z\) is a normalizing constant, also known as the marginal likelihood or evidence.
From this parameterization, the posterior expectation can be computed by setting \(e^{-z^2} = u_q\) and \(z=z_q\), where for \(q = 1, \ldots, Q\), \(z_q\) is the \(q\)th Gauss-Hermite quadrature node, and \(u_q\) is the \(q\)th Gauss-Hermite weight.

\subsection{Marginalized Adaptive Beamformer}\label{section:beamformer}
\textit{
Marginalized Adaptive Beamformer.
}~
Given the posterior of \(c\), we compute the marginalized MVDR beamformer as
{
\begin{align}
   y\left(p\right) 
   &= \int \vw^{\dagger}\left(p,c\right) \, \vx^{\rm{d}}\left(p, c\right) \, \pi_p\left(c \mid \vx_{1\text{:}K}\right) \, \mathrm{d}c \\
   &\approx \frac{1}{\sum^{Q}_{q=1} v_q} \sum^{Q}_{q=1} 
   v_{q} \, \vw^{\dagger}\left(p,c\right) \, \vx^{\rm{d}}\left(p, c\right),
   \label{eq:focal_point_intensity}
\end{align}
}%
where the unnormalized posterior density is 
{%
\begin{align}
   v_{q} = u_q \, \ell_p\left(
   \vx_{1\text{:}K} \mid 
   \sqrt{2} \, z_q \, \sigma_c + \mu_c\right),\label{eq:importance_weight}
\end{align}
}%
and \(\vw\left(p, c\right)\) are set as the MVDR fading weights.
As mentioned in \cref{section:signal_model}, the steering vector is set as \(\va = \mathbf{1}\).
Also, we reuse the covariance \(\widehat{\mSigma}_{\vx,\,\text{DL}}\) used to compute the likelihood in~\cref{eq:importance_weight} for finding the MVDR weights in~\cref{eq:mvdr}, \textcolor{black}{where the noise covariance estimate is set as \(\widehat{\mSigma} = \widehat{\mSigma}_{\vx,\,\text{DL}}\)}.

\textit{
Imaging with Marginal Likelihoods.
}~
A unique aspect of our Bayesian scheme is that we can obtain the \textit{evidence} of the data.
That is, for a pixel \(p\), we can estimate the marginal likelihood
\begin{align}
    Z\left(p\right) 
    =
    \int \widetilde{\ell}_p\left(\vx_{1\text{:}K} \mid c\right) \rho\left(c\right)  \mathrm{d}c
    \approx
    \sum^{Q}_{q=1} v_q.
    \label{eq:marignal_likelihood}
\end{align}
This represents how likely it is to observe the data acquired on \(p\) given our model of the system, in particular, the prior.
On the flip side, the marginal likelihood can also be interpreted as the evidence for our \textit{model}.
If our goal is to detect signals coherent with the \cref{eq:marignal_likelihood} and the prior over the \acrshort*{sos}, this means that we can treat this as the probability of detecting such signal.
Therefore, we can use the marginal likelihood in \cref{eq:marignal_likelihood} to image signals.
We will refer to this method as ``ML'' in the experimental section.
Also, we will treat the expression in \cref{eq:marignal_likelihood} as the natural scale of this imaging signal, where visualization is done in log scale (``nats''), similarly to other power-based imaging signals.

\textit{
Computational Cost.
}~
The computational cost of computing the marginalized beamformer is proportional to the number of quadrature points \(Q\).
More precisely, evaluating a single quadrature point is equivalent to computing a single MVDR beamformer, where the computation linearly increases in \(Q\).


\setcounter{figure}{1}

\hspace{-0.25in}
\begin{minipage}{0.29\textwidth}
\centering
\begin{threeparttable}
\small
\caption{Simulation Settings}\label{table:simulation}
\setlength\tabcolsep{1pt}
\begin{tabular}{lrl}
\toprule
    Parameter & Val. & Unit \\ 
\midrule
    Bathymetry depth  & 100  & {\small\texttt{[m]}} \\
    Number of sensors    & 30   & \\
    Length of array      & 1   & {\small\texttt{[m]}} \\
    LFM\; center freq. & 30k & {\small\texttt{[Hz]}} \\
    LFM\; bandwidth        & 20k & {\small\texttt{[Hz]}} \\
    LFM\; duration         & 50\(\mu\)  & {\small\texttt{[s]}} \\
    Ambient noise power  & 80\tnote{a} & {\small\texttt{[dB\tnote{b}\;]}} \\
    Signal power  & 190  & {\small\texttt{[dB\tnote{b}\;]}} \\
    Sampling freq. & 500k & {\small\texttt{[Hz]}} \\
    Sampling duration   &  0.3 & {\small\texttt{[s]}} \\
\bottomrule
\end{tabular}
\begin{tablenotes}
\item[a] See the text in \cref{section:simulation_setup}.
\item[b] unit: re 1 \(\mu\)Pa at 1m.
\end{tablenotes}
\end{threeparttable}
    \vspace{-3ex}
\end{minipage}%
\begin{minipage}{0.39\columnwidth}
    \vspace{3ex}
    \includegraphics[scale=0.9]{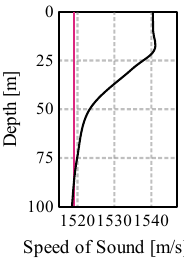}
    \captionof{figure}{\acrshort*{sos} profile (black). The red line (1519 m/s) is the value used by DAS and MVDR and the mean of our prior on \(c\).}\label{fig:bathy}
    \vspace{-3ex}
\end{minipage}

\setcounter{figure}{2}

\section{Evaluation}\label{section:experiments}
\subsection{Sonar Simulation Setup}\label{section:simulation_setup}

%
\textit{
Simulation Implementation.
}~
We apply our beamformer to active sonar imaging.
We simulate point targets with the Bellhop propagation model~\cite{porter_gaussian_1987,bucker_simple_1994}, with an implementation similar to Larsson and Gillard~\cite{larsson_waveform_2004}, where only a single ``ping'' is used.
The simulation settings are shown in~\cref{table:simulation}, while the \acrshort*{sos} profile is shown in~\cref{fig:bathy}.
\textcolor{black}{
For the ambient noise, we combine colored noise and white noise, which is more realistic for ocean acoustics~\citep[Fig. 8.13]{ainslie_principles_2010}.
The colored noise follows a power law with exponent 1.4, where the power at 1 Hz is approximately set to be 85 dB re 1 \(\mu\)Pa at 1m.
The white additive Gaussian noise has a power 80 dB re 1 \(\mu\)Pa at 1m, which represents a relatively high level of ambient noise encountered on a rainy day~\citep[Fig. 8.13]{ainslie_principles_2010}.
The resulting peak SNR is roughly 40 dB.
}
The bathymetry is set to be flat.
The array and the linear frequency-modulated (LFM) source are at a depth of 70 m, while the targets are at a depth of 90 m.
The beam resolution is evaluated with a single point target at a range of 36 m, while 5 point targets, which are 1 m apart, are arranged as a ``cross'' for qualitative evaluation.

After simulating propagation, the signal is quantized into 16-bit resolution.
Then, we apply a time-varying gain of \(G_{\mathrm{TVG}}\left(t\right) = 20 \log r \approx 20 \log \left(\pi t \, c\right) \) \texttt{[dB]}, and demodulate the signal with a quadrature demodulator with a decimation ratio of 4.
Range compression is then done by matched filtering.

\textit{
Baselines and Algorithm Setup.
}~
For the baselines, we use the DAS beamformer and the MVDR adaptive beamformer~\cite{frost_algorithm_1972}.
Both use a constant \acrshort*{sos} of 1519 m/s.
The DAS beamformer uses Hann fading weights.
For the MVDR beamformer, we use subarray averaging, diagonal loading, and forward-backward averaging.
We use \(\textcolor{black}{K} = N / 2\) for both the MVDR beamformer and our Bayesian variant (Bayes MVDR).
We configure our Bayesian approaches (Bayes MVDR and ML) to use \(\text{SNR}_0 = 15\) dB, \(\mu_c = 1519\) and \(\sigma_c = 0.3\).
\textcolor{black}{
For the number of quadrature points, \(Q \in \{8, 16, 32\}\) is considered.
(Note that our method, Bayes MVDR, is the only method that uses numerical quadrature. Therefore, other methods do not involve \(Q\).)
}
\textcolor{black}{
Each beamformer outputs a 2-dimensional image of the response across the azimuthal and range direction, which are perpendicular and parallel to the direction the array faces.
}

\begin{figure}[t]
    \centering
    \hspace{-2em}
    \subfloat[Mean]{
        \includegraphics[scale=0.85]{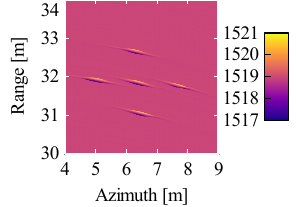}\label{fig:sos_mean}
    }
    \subfloat[Variance]{
        \includegraphics[scale=0.85]{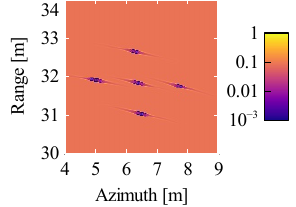}\label{fig:sos_var}
    }
    \caption{
    Posterior estimates of the \acrshort*{sos} with \(Q=8\).
    }\label{fig:sos}
    \vspace{-3ex}
\end{figure}

\begin{table}[t]
\caption{Resolution and Multipath Artifact Suppression}
\begin{center}
\begin{threeparttable}
\begin{tabular}{lrrrr}
    \toprule
     & \multicolumn{1}{c}{\multirow{1}{*}{DAS}} & \multicolumn{1}{c}{\multirow{1}{*}{MVDR}} & \multicolumn{1}{c}{Bayes MVDR} & \multicolumn{1}{c}{ML}
     \\ 
     \midrule
      FWHM\tnote{1} \texttt{[m]}  & 2.27  & 0.31   & 0.27    & 0.08 \\
      RPMAL\tnote{2} \texttt{[dB]} & -9.96 & -9.44  & -23.99  & -42.01 \\ \bottomrule
\end{tabular}
\begin{tablenotes}
\item[1] Full width at half maximum \item[2] Relative peak multipath artifact level
\end{tablenotes}
\end{threeparttable}
\end{center}
\label{table:performance}
\vspace{-3ex}
\end{table}
\begin{figure}[h]
    \centering
    \input{plots/azim_resol_plot}
    \caption{Normalized azimuth resolution plot for a point target at a range of 30 m. 
    The dotted line shows the azimuth position of the point target.}\label{fig:quantatitive}
    \vspace{-4ex}
\end{figure}

\vspace{-1ex}
\subsection{Simulation Results}

\textit{
Qualitative Results.
}~
The qualitative results are shown in~\cref{fig:qualitative}.
The 5-point targets forming a ``cross'' are near the range of 32 m, while multipath artifacts appear near the range of 39 m.
In terms of resolution, our marginalized beamformer (Bayes MVDR) results in lower sidelobes compared to the DAS and MVDR beamformer, resulting in better visibility of the targets.
Also, it suppresses the multipath artifacts more effectively compared to the other baselines.
Meanwhile, ML imaging performs even better than Bayes MVDR, with the highest resolution and barely visible multipath artifacts. 
Note that, for the ML image, we did not restrict the dynamic range whatsoever. 
Therefore, ML clearly achieves much stronger multipath artifact suppression.

In the second row of~\cref{fig:qualitative}, we evaluate the effect of the number of quadrature points, which is directly connected with the computational cost of our method.
As shown, the differences between \(Q = 8, 16, 32\) are marginal.
Quantitatively, the root mean squared error between \(Q=8\) and \(Q=32\) is only -39.13 dB. 
Therefore, the benefits of our beamformer can be enjoyed with a relatively small number of quadrature points.

To understand the behavior of the marginalized beamformer (Bayes MVDR), we visualize the posterior estimates of the \acrshort*{sos} in \cref{fig:sos}.
In \cref{fig:sos_mean}, we can see that the posterior mean of the \acrshort*{sos} deviates from the prior mean near the point targets.
Intuitively, the posterior of the \acrshort*{sos} is steering the beam towards the direction of the signal.
At the same time, the degree of steering is regularized by the \textcolor{black}{uncertainty of the posterior, which can be quantified through the posterior variance shown in \cref{fig:sos_var}}.
That is, when the uncertainty is high, the signal is averaged over a wider range of \acrshort*{sos}, smoothing out the signal.
This smoothing effect is what suppresses the sidelobes and the multipath artifacts.

\textit{
Quantitative Results.
}~
We quantitatively evaluate the resolution and the artifact suppression level.
The resolution is measured over a 1-dimensional cutaway along the azimuth axis of one of the point targets, where we normalize the signal level to be in the range of \([0,1]\) and compute the full width at half-maximum (FWHM) in meters.
For the artifact suppression level, we compute the relative peak multipath artifact level (RPMAL), which is the ratio of the peak signal level over the region where the true targets are versus where they are not.
For both ML and Bayes MVDR, we used \(Q = 8\).
The objective metrics are organized in \cref{table:performance}.
Bayes MVDR can be seen to achieve better resolution compared to the DAS and MVDR beamformers, where the multipath artifact level is 14dB lower than baselines.
Furthermore, ML imaging achieves twice as accurate resolution and close to no artifacts.
We visualize the full azimuth resolution plot in~\cref{fig:quantatitive}.

\vspace{-1ex}
\section{Conclusions}\label{section:conclusions}
In this work, we have presented a Bayesian adaptive beamformer for imaging that marginalizes away the speed of sound.
On simulated active sonar data, our beamformer achieved better resolution and improved suppression of multipath artifacts relative to standard DAS and MVDR beamformers.
In practice, especially for sonar imaging, it is common to compound the data acquired from multiple ``pings'' in a synthetic aperture scheme~\citep{austeng_use_2013,hayes_synthetic_2009}.
Combined with this, the resulting images would result in even better resolution.
Furthermore, we demonstrated that the marginal likelihood (ML) of the model can be used as a signal for imaging.
ML imaging showed promising results and warrants further investigation in the future. Indeed, our method should be applicable to any array-based imaging method, including sonar imaging, medical ultrasound, and seismic imaging: it would be interesting to explore applications in each such context.
In the future, evaluating the proposed method on real data acquired from a moving platform and incorporating various uncertainties in the signal processing chain, not just the speed of sound as we considered here, into a Bayesian model would be worth investigating.
Lastly, relaxing the Gaussian noise assumption to other noise distributions could further improve the performance.


\bibliographystyle{IEEEtran}
\bibliography{bstcontrol,references}


\end{document}

%% file: krkmath.tex
\usepackage{amsmath,amssymb,amsthm}
\usepackage{mathtools}

\usepackage{multirow}
\usepackage{nicematrix}

\usepackage{dsfont}

\usepackage{proof-at-the-end}

\usepackage{mdframed}
\newmdtheoremenv{framedtheorem}{\textbf{Theorem}}
\newmdtheoremenv{framedproposition}{\textbf{Proposition}}
\newmdtheoremenv{framedlemma}{\textbf{Lemma}}

\usepackage{url}
\usepackage{hyperref}
\hypersetup{
    colorlinks = false,
}
\usepackage[capitalise, nameinlink]{cleveref}

\crefname{framedtheorem}{Theorem}{Theorems}
\crefname{framedproposition}{Proposition}{Propositions}
\crefname{framedlemma}{Lemma}{Lemmas}

\pgfkeys{/prAtEnd/global custom defaults/.style={
    end,
    restate,
    text link={\textit{Proof.} See the \hyperref[proof:prAtEnd\pratendcountercurrent]{\textit{full proof}} in~\cref{section:proofs}.
    }
  }
}

\newcommand*\xbar[1]{%
  \hbox{%
    \vbox{%
      \hrule height 0.6pt 
      \kern0.33ex
      \hbox{%
        \kern-0.1em
        \ensuremath{#1}%
        \kern-0.1em
      }%
    }%
  }%
}

\newcommand{\abs}[1]{{\left|\,#1\,\right|}}

\newcommand{\symbfup}[1]{\mathbf{#1}}

\newcommand{\va}{\symbfup{a}}

\newcommand{\vc}{\symbfup{c}}

\newcommand{\vn}{\symbfup{n}}

\newcommand{\vw}{\symbfup{w}}
\newcommand{\vx}{\symbfup{x}}

\newcommand{\mI}{\symbfup{I}}
\newcommand{\mJ}{\symbfup{J}}

\newcommand{\mSigma}{\symbfup{\Sigma}}

%% file: plots/azim_resol_plot.tex
\begin{tikzpicture}
	\begin{semilogyaxis}[
		xlabel={\footnotesize%
		    Azimuth [m]
		},
		ylabel={\footnotesize%
		    Relative Power [dB]
		},
		log basis y=10,
        yticklabel={\pgfmathparse{20*(\tick)}\pgfmathprintnumber[fixed]{\pgfmathresult}},
	    xtick ={-4, -3, -2, -1, 0, 1, 2, 3, 4},
	    ytick ={0.001, 0.01, 0.1, 1},
	    axis line style = thick,
	    xticklabel style={
            /pgf/number format/fixed,
            /pgf/number format/precision=2
        },      
        minor tick length=1.5pt,
        major tick length=2.5pt,
        every tick/.style={
            black,
            semithick,
        },
	    xmin  =-4,
	    xmax  =4,
	    ymin  =0.001,
	    ymax  =1,
	    scale = 1,
	    width =7cm,
	    height=4.cm,
	    legend style = { 
	        legend columns = 2,
	        draw           = none,
	        at={(0,1)},
            anchor=south west,
            legend cell align={left},
	    }]
	]
	\input{plots/das_azim_resol}
	\addlegendentry{\footnotesize%
	    DAS
	}
	\input{plots/mv_azim_resol}
	\addlegendentry{\footnotesize%
	    MVDR
	}
 
	\input{plots/bayes_8_azim_resol}
	\addlegendentry{\footnotesize%
	    Bayes MVDR\;
	}

	\input{plots/lml_8_azim_resol}
	\addlegendentry{\footnotesize%
	    ML\;
	}
 
	\addplot[thick, densely dashed, color=gray] coordinates {
	    (0, 0.00001)
	    (0, 1)
	};
	\end{semilogyaxis}
\end{tikzpicture}